\RequirePackage{lineno} 
\documentclass[aps,prl,onecolumn,notitlepage,fontsize=12pt]{revtex4-1}
\usepackage{amsmath}
\usepackage{amssymb}
\usepackage{graphicx}
\usepackage{color}
\usepackage{float} 
\usepackage{geometry}
\usepackage{graphicx}
\usepackage{sidecap}

\pdfoutput=1

\newcommand{\eq}{\begin{equation}}
\newcommand{\eeq}{\end{equation}}

\begin{document}

\title{Rapid production of defect-free beryllium ion Coulomb crystals}

\author{Qiming Wu}
 \email{qiming.wu@nyu.edu}

\author{Melina Filzinger}%
\author{Yue Shi}

\author{Jiehang Zhang}
\affiliation{%
Physics Department, New York University, New York, USA
}%

\author{Zhihui Wang}
\affiliation{%
State Key Laboratory of Quantum Optics and Quantum Optics Devices, Shanxi University, China
}%

\date{\today}

\begin{abstract}
Trapped atomic ions find wide applications ranging from precision measurement to quantum information science and quantum computing.  Among the different atomic species employed, beryllium ions are widely used due to its light mass and convenient atomic structure. However, the ion loading process requires a high temperature for
sufficient vapor pressure, generating undesirable gas load for the background vacuum and limiting the lifetime of a long ion chain. Here,  we demonstrate a simple method to rapidly produce pure linear chains of beryllium ions with pulsed laser ablation, serving as a starting point for large-scale quantum information processing. Our method is fast compared to thermal ovens; reduces the vacuum load to only 10$^{-12}$ Torr level; yields a short recovery time of a few seconds; and also eliminates the need for a deep ultraviolet laser for photo-ionization. In addition, we apply feedback control to obtain defect-free ion arrays with desirable lengths.  

\end{abstract}

\maketitle


\section{\label{sec:level1}Introduction}

Atomic ions confined in electromagnetic traps have been at the fore-front of accurate optical atomic clocks~\cite{Brewer2019}, fundamental symmetry searches~\cite{Cairncross2017}, prototype quantum computers~\cite{Debnath2016}, and large-scale quantum simulators for many-body physics~\cite{Zhang2017}. The deep confining potential offers tight controls in a well-isolated environment, and precisely tailored laser pulses can be used to either probe fundamental physics ~\cite{sanner2019optical,Cairncross2017} or drive high-fidelity entangling gates~\cite{Gaebler2016}. Among different trapping techniques, the widely-used linear Paul trap employs fast oscillating radio-frequency electric fields to generate a net transverse confinement, in combination with axial static fields for three-dimensional trapping~\cite{Paul1990}. The trap is anisotropic with an aspect ratio is tuned such that ions self-assembled into a linear Coulomb crystal with minimum RF-induced micro-motion. Through dispersively driving the collective quantum harmonic oscillator modes, the internal electronic states of ions can be coupled to realize either Ising-type interaction~\cite{Zhang2017} or two-qubit gates~\cite{Debnath2016}. However, this same Coulomb interaction stores enough classical potential energy, such that a collision from residual background gas components could `melt' the ion chain, even in the presence of ultra-high vacuum of 10$^{-11}$ Torr pressure~\cite{Zhang2017}, leading to extensive efforts to build cryogenic ion traps with considerable complexity overheads~\cite{Poitzsch1996, Pagano2018}. Conventional ion loading from a thermal oven exerts a high vapor pressure burden, which creates a run-away problem since the contaminated vacuum further reduces the ion chain lifetime, contributing as a critical limitation for scaling up trapped-ion quantum information processors~\cite{Zhang2017}. A rapid loading scheme with a small background gas load is hence highly desirable for state-of-the-art experiments. 

\begin{figure*}
    \centering
    \includegraphics[width=0.95\textwidth]{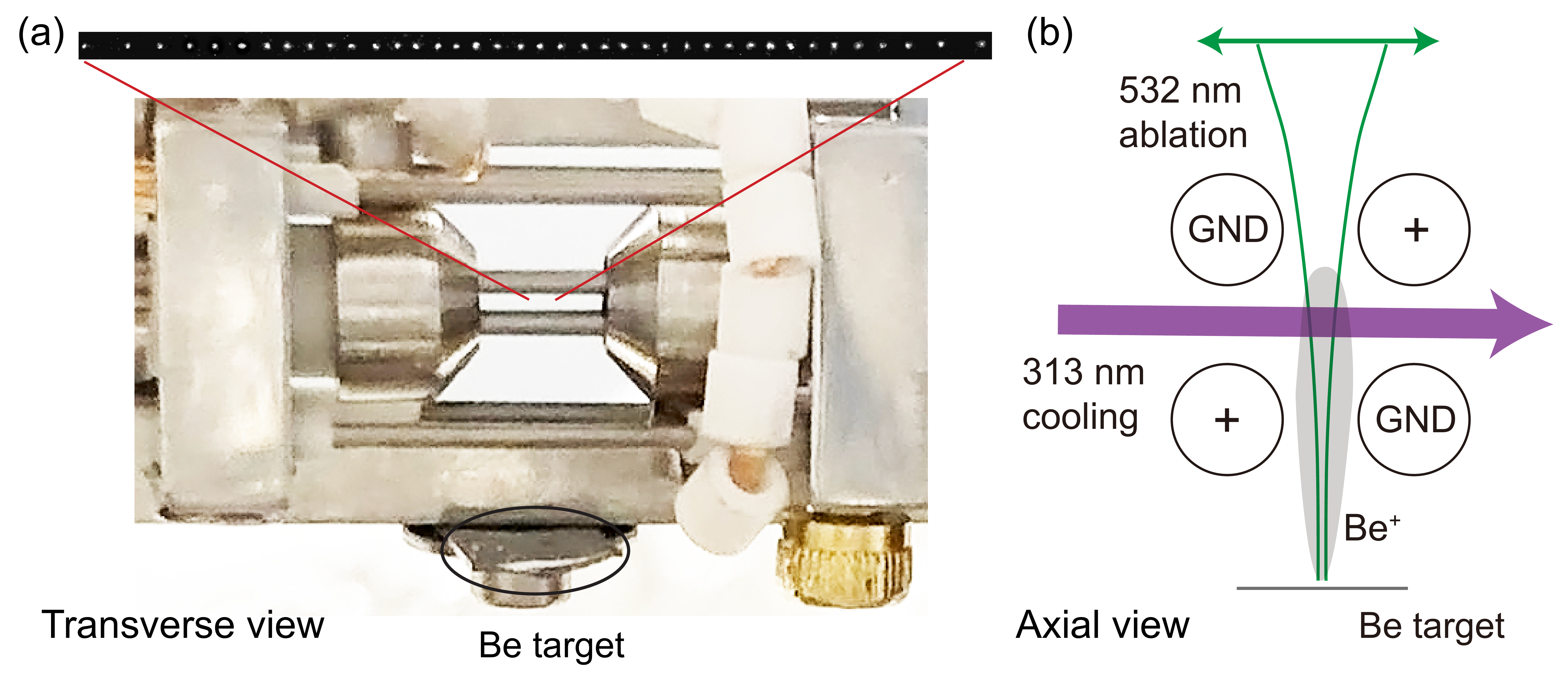}
    
    \caption{Ion trap set up and laser geometries. (a) Picture of the 4-rod trap inside the vacuum chamber, view from the transverse direction.  Inset shows 39 Be$^{+}$ ions trapped in a linear chain. (b) Illustration of ablation and cooling laser geometry, viewed from the axial direction. A pulsed laser at 532 nm is focused down to the target after passing through the trap electrodes, so that the ablated plasma plume (grey ellipse) is directed into the trap. A stable laser at 313 nm (purple arrow) cools the beryllium ions on the first dipole-allowed transition.}
    \label{fig:Trap set-up}
\end{figure*}

The ion we choose to work with is singly-charged beryllium ($^{9}$Be$^{+}$), where the simple level structure allows efficient laser cooling; its light mass enables fast quantum gates; and the ground-state hyperfine structure allows for convenient qubit manipulation and detection. These properties have indeed enabled quantum gates with the highest fidelity among any platform~\cite{Gaebler2016}. However, loading of beryllium ions is conventionally carried out with a thermal atomic source heated to high-temperatures at a fraction of the melting point of 1560\,K, to produce a diffusive neutral atom beam, followed by either electron-impact ionization or photo-ionization with a deep ultraviolet (DUV) laser~\cite{Wolf2018}. Besides the gas load, this loading process can be time-consuming because of the long time-constant at the minute scale~\cite{Tan2016}. Moreover, the consecutive ionization methods are also invasive: the electron gun produces ions with high kinetic energies that are hard to cool and creates patch charges that perturb the trap electric fields~\cite{Leibrandt2007}.  Alternatively, photo-ionization is a cleaner process, but for Be the DUV photon energies at 235 nm or 266 nm~\cite{Wolf2018} are above the band-gaps of most dielectric materials, thus presenting a significant risk of photo-electric charging~\cite{Harlander2010}. These undesirable effects from electrons and UV photons could readily affect ion motion at the single-quanta level, hampering high-fidelity entangling gates.

A different method is to use laser ablation loading, which has been successfully demonstrated with many different ion species for RF traps~\cite{Leibrandt2007,Hendricks2007,Sheridan2011,Zimmermann2012,Olmschenk2017,Shao2018,Vrijsen2019,Hahn2019,Sameed2020}. We demonstrate, for the first time to our knowledge, a reliable scheme for producing ion Coulomb crystals of $^{9}$Be$^{+}$ with direct laser ablation in the plasma regime. Our scheme has the advantage of ultra-low vacuum pressure overload, achieved by directly loading ions from the plasma plume induced by focusing a visible turn-key Nd:YAG laser. While many recent ablation techniques operate in the neutral regime followed by photo-ionization for isotope selectivity~\cite{Hendricks2007,Sheridan2011,Shao2018,Vrijsen2019}, we obtain entropy-free crystals of beryllium ions because of the presence of only one stable isotope ($^{9}$Be) and the clean chemical composition of the target. Our technique contributes to a toolbox to enable large-scale quantum processors beyond 100 ions in a compact room temperature apparatus.

\section{Experimental Setup}

We use a linear Paul trap in a four-rod configuration similar to Ref.~\cite{Poitzsch1996}, shown in Fig.~\ref{fig:Trap set-up}(a). This trap is capable of holding hundreds of beryllium ions in a linear chain, and up to 70 ions at high enough secular frequencies suitable for quantum logic operations. We construct the trap using four cylindrical tungsten rods with diameters of 500 $\mu m$, placed at the corners of a square with center-to-center distance of 950 $\mu m$ (Fig.~\ref{fig:Trap set-up}(b)), insulated with alumina ceramic disks with clearances holes matching the size of the rods. A pair of endcaps machined from stainless steel is placed in the axial direction with a spacing of 4.6 mm, sleeved outside of the ceramic disks. The conical shapes of the endcaps facing the ions provide an enhanced electric field strength and in turn a higher axial trapping potential. We typically apply an RF drive at $\Omega = 2\pi\times$ 71.8 MHz with 33 dBm of power and a DC voltage of 250 V on the endcaps, corresponding to radial and axial trap frequencies of $\{\omega_{x,y}, \omega_z\} = 2\pi\times\{5.0, 0.45\}$ MHz. The ablation target is a 10 mm$\times$ 20 mm $\times$ 0.25 mm rectangular beryllium foil (99.8+\% purity, Sigma-Aldrich) mounted on the trap holder under the RF electrodes, at a distance of d = 8.3 mm from the trap center (Fig.~\ref{fig:Trap set-up}(a)). 

\begin{figure*}
    \centering
    \includegraphics[width=1.0\textwidth]{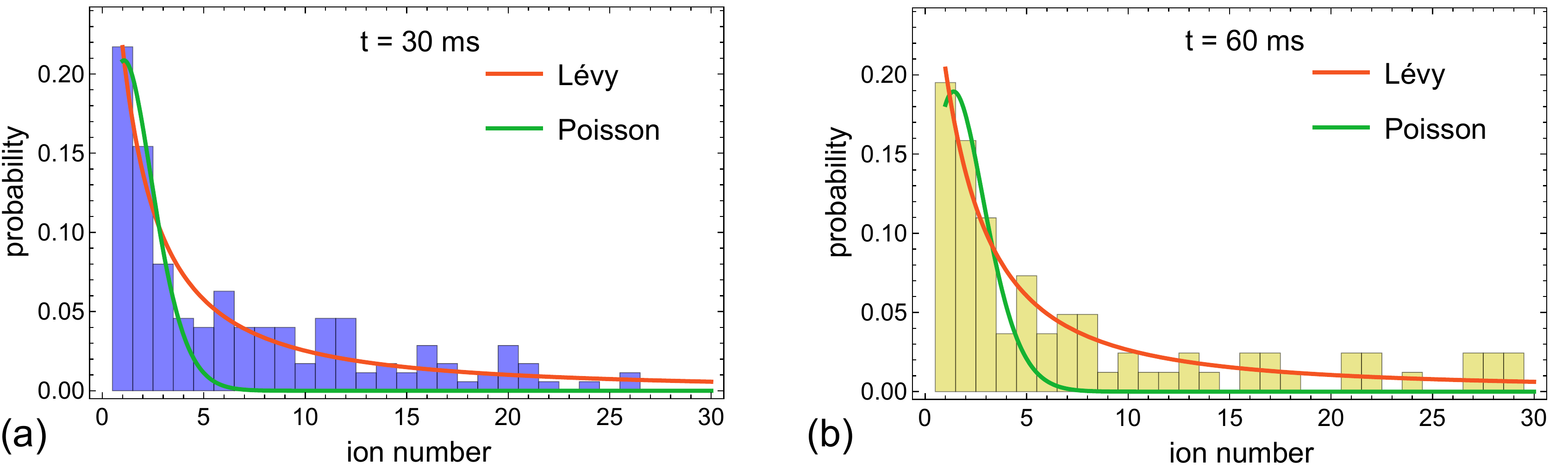}
    \caption{Histogram of ion number distribution with 100 loading trials, for different ablation durations. (a) Ablation time t $=$ 30 ms. (b)  t $=$ 60 ms. Both histograms are fitted to Poisson (green line) and Lévy (red line) distributions. The Poisson fits fails to capture the long-tails, whereas the Lévy fits give much better agreements. The loaded ion number is not directly proportional to the duration, which together with the tails indicates that the ion number is dominated by crystallization dynamics, rather than the ion number from the ablation plasma.}
    \label{fig:Hiostagram}
\end{figure*}

We use a Q-switched Nd: YAG laser at 532 nm (EL-532-1.5W, CNI lasers) that outputs nanosecond pulses with controllable repetition rates and pulse energies. Although there is 1 W of green light at our disposal, we find it usually sufficient to operate the laser slightly above the threshold (7 A of pump current). After an acoustic-optical modulator (AOM) and beam-shaping optics, the power directed into the trap is 80 mW at a repetition rate of 4 kHz. With a pulse duration of 2.2 ns and single pulse energy of about 20 $\mu$J, we can load a chain of up to 20 ions in 30 ms (120 pulses).  Measuring with an extractor gauge (IE514, Leybold), we observe a maximum vacuum spike of $4\times10^{-12}$ Torr, which rises faster than the gauge update rate of about 1\,s and decays to the base pressure in less than ten seconds, a time-constant faster than that of specially designed thermal ovens~\cite{Ballance2018, Gao2020}. The ablation and cooling laser geometries are illustrated in Fig.~\ref{fig:Trap set-up} (b). After an AOM (not shown in the diagram) to control the number of pulses and the energy per pulse, we focus the 532\, nm laser beam down to a 1/$e^2$ diameter of 60 $\mathrm{\mu m}$ on the beryllium target. Ions from the slow tail of the kinetic energy distribution are trapped by the RF potential and laser-cooled to a Coulomb crystal. The ablation laser fluence of 1.5\,$\mathrm{J/cm^2}$ at the beam center is in good agreement with the threshold measured in Ref.~\cite{Sameed2020}. Such pulse energies only slightly above the threshold should produce ions with kinetic energies less than 10 eV~\cite{Sameed2020}. For diagnostic purposes, we use a commercial long working distance microscope to observe white light from the laser-induced plasma on the beryllium target, facilitating the initial alignment as well as recoveries after changes in the experimental setup. We use a frequency quadrupled laser at 313 nm (FHG Pro, Toptica) to address the $^{2}S_{1/2}$ to $^{2}P_{3/2}$ transition in Be$^{+}$ and collect fluorescence either on a photo-multiplier tube (PMT) or an intensified CMOS camera. 

\begin{figure}[hb!]
    \centering
    \includegraphics[width=0.6\textwidth]{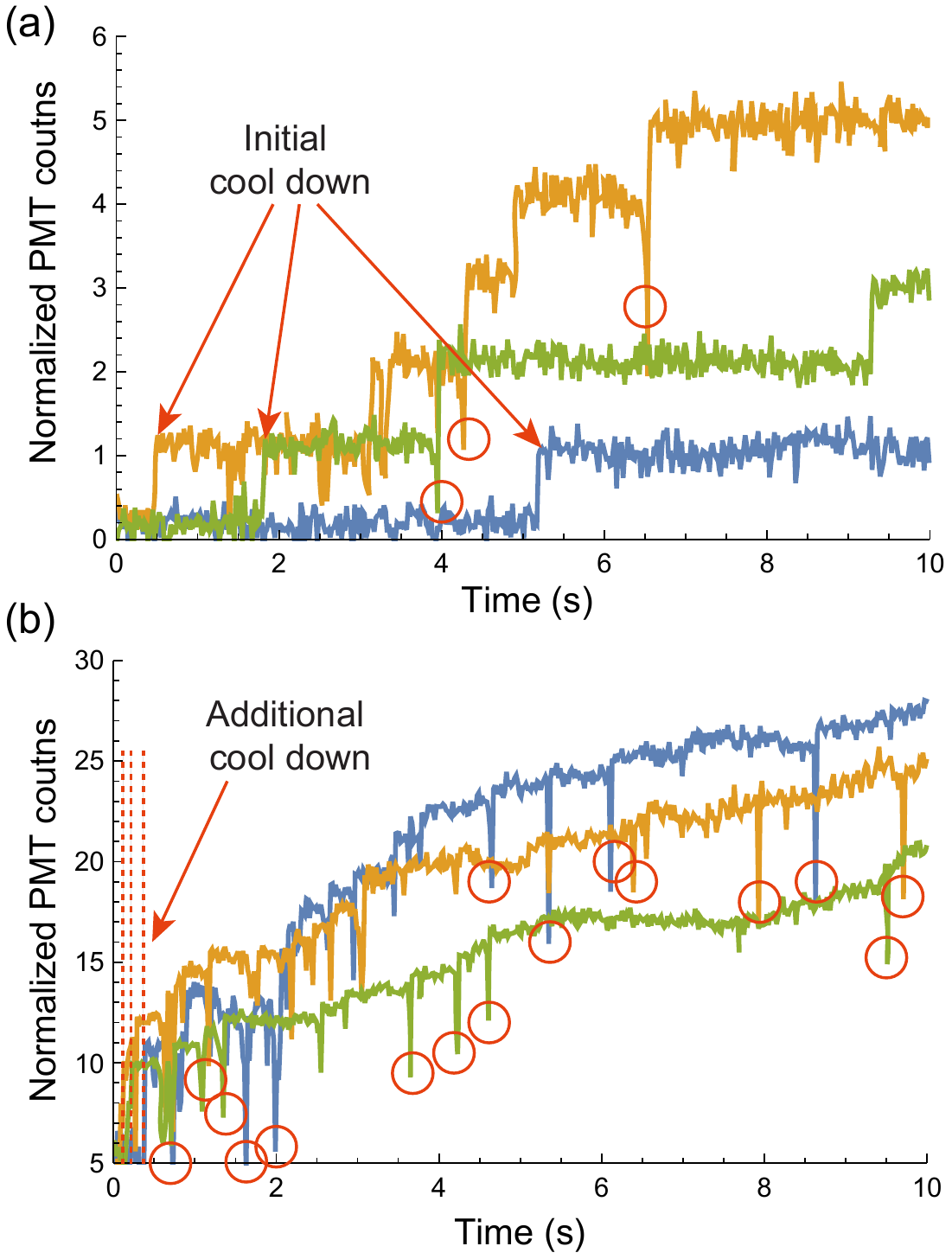}
    \caption{Crystallization dynamics of ablation loading. PMT counts are normalized with repect to photon counts per ion. (a) Initial loading from an empty trap. Red arrows show initial cool down time of 0.5 s, 1.8 s, 5.2 s, respectively.  (b) Crystallization dynamics with five ions pre-loaded into the trap. Red dash line shows additional crystallization time of 0.1 s, 0.2 s, 0.4 s, respectively. Red circles show instantaneous melting induced by a new ion before re-crystallization.}
    \label{fig:crystalization}
\end{figure}

\section{Statistics and feedback control}

\begin{figure*}
    \centering
    \includegraphics[width=0.9\textwidth]{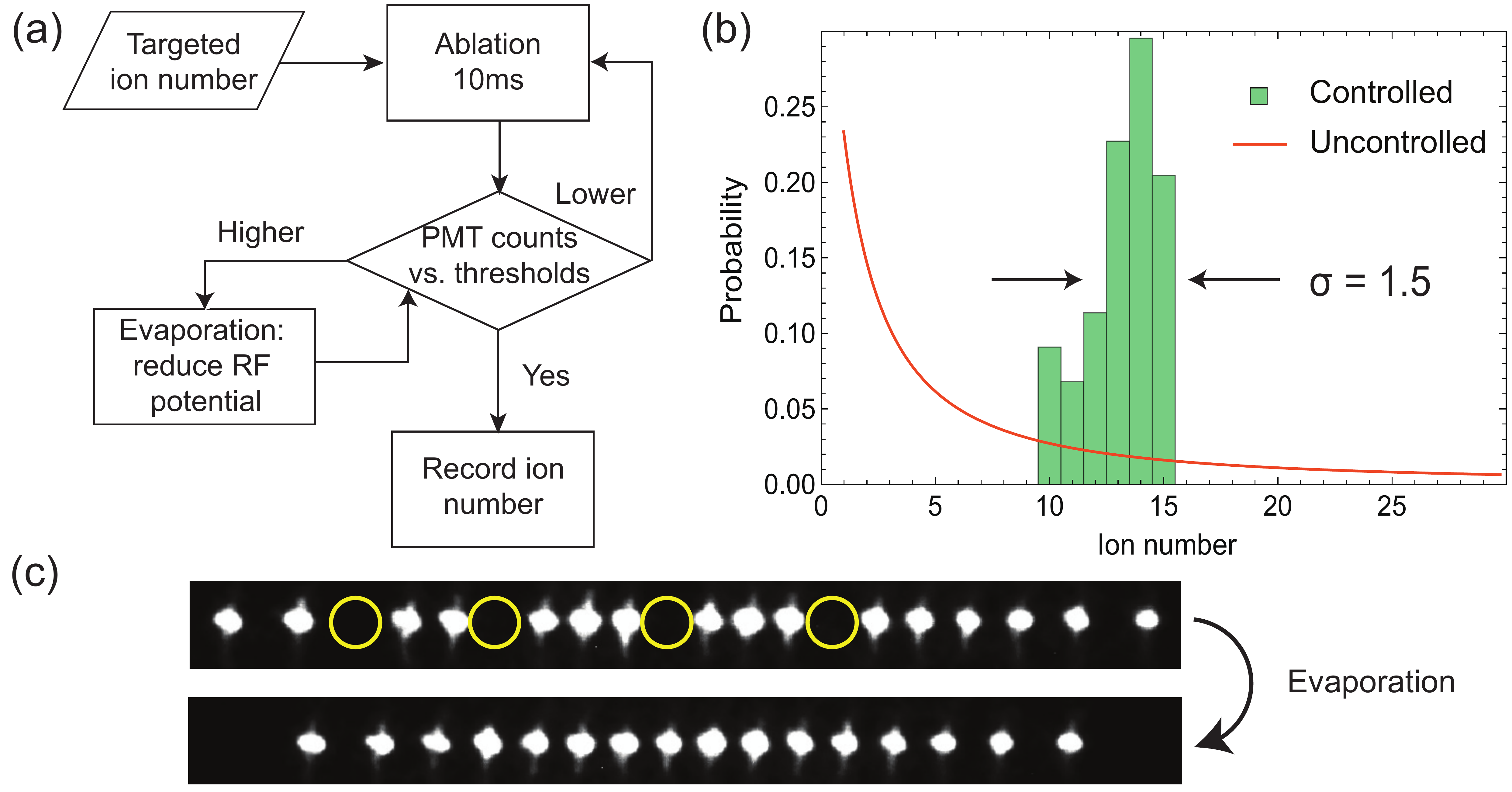}
    \caption{Feedback control for entropy-free chains with desirable ion number. (a) Block diagram of the control sequence for loading with conditional logic. Upon an input ion number, we count the ion fluorescence photons with a PMT to determine whether the existing ion number is within the targeted range. Feedback is applied either to add by extra ablation or subtract via evaporation: lowering the RF potential and reduce cooling for hot ions to leave the trap. (b) Statistic of ion number after controlled feedback (green bars). (c) Purifying the ion chain with evaporation. A previous fit-curve (red solid line) without the control sequence is shown for comparison. The standard deviation of the ion number is reduced to 1.5. (c) Dark ions of BeH$^{+}$ (yellow circles) formed after inelastic background gas collisions can be spilled out of the trap by applying the same evaporation method for a shorter duration, leaving only bright atomic ions.}
    \label{fig:Control loading}
\end{figure*}

The intrinsic randomness of the loading process, whether from thermal or ablation-plasma sources, gives rise to  distributions obeying Poisson statistics in the absence of correlated dynamics. However, whether the number of Coulomb-crystallized ions follows Poisson distribution has never been investigated to the best of our knowledge. We sample the ion number distribution by counting photons with repeated measurements on a 1 ms long Doppler cooling sequence. By collecting about 3500 photon counts per ion, we can distinguish the difference from a single ion in the presence of about 30 ions with a signal above the photon shot noise.
Fig.~\ref{fig:Hiostagram} shows the histogram of the sampled ion number distribution from ablation loading. The Doppler cooling laser is set to five times of saturation power, $-$10 MHz ($-\Gamma/2$) detuned from the $S_{1/2} - P_{3/2}$ transition of Be$^{+}$.  To enhance the cooling of fast ions, we apply two additional frequencies with detunings of $-$120 MHz and $-$1.2 GHz. With identical laser cooling trap parameters, we apply two different ablation durations, 30 ms (Fig. 2(a)) and 60 ms (Fig. 2 (b)), respectively. Both statistics are characterized with a long tail, and fitting to Poisson distributions fails to capture this. We in turn fit the data to a Lévy distribution and find a significantly better agreement. Such statistics can hardly be explained with a laser cooling theory of individual non-interacting particles~\cite{Wineland1979}, and calls for an understanding of the crystallization dynamics in the presence of the Floquet drive from the RF and the long-range Coulomb interactions~\cite{Blumel1989}. 

While an extensive study of this classical many-body dynamics is currently under investigation in our group and beyond the scope of this paper, we point out in Floquet-driven systems more than two particles readily give rise to chaotic dynamics~\cite{Blumel1989}. Theory of such dynamics is not only difficult because of the lack of analytical methods, but also challenging in numerics because of the long timescales (10 seconds) and the fast RF oscillations (13 ns per cycle) involved. Nevertheless, we provide a plausible explanation for the deviation from Poisson statistics: if a small ion crystal is already formed, hot ions in higher orbits in the trap can experience sympathetic cooling, leading to the enhanced formation of longer ion strings. 

To test our hypothesis, we use identical loading sequences with and without an existing ion chain and measure the time constant for ions to crystallize. Specifically, we apply ablation pulses of 10 ms, and record PMT counts in the following 10 s, capturing the time it takes for an ion chain to crystallize. Fig.~\ref{fig:crystalization} (a) shows crystallization dynamics of three ablation loading trials with an empty trap. The red arrows show initial cool down time of 0.5\,s, 1.8\,s, 5.2\,s, when the first ion enters the trap. Then more ions crystallize successively, as indicated by the step-like increases of PMT counts.  For comparison, we repeat the same sequence but with five ions pre-loaded into the trap. We observe a rapid appearance of about five additional ions joining the crystal, with cooling times of 0.1\,s, 0.2\,s, and 0.4\,s for three different trials, respectively. Then more ions come in sequentially until a steady chain is formed. Note that the new comings ions would first `melt' the crystal in a transient, then recrystallize and form a longer chain. This results in the dips before the stepped increases of PMT counts, marked by the red circles in Fig.~\ref{fig:crystalization}. Compared to an empty trap, with preexisting ion chains we observe the initial cool down to be ten-fold faster, with higher final numbers and a similar steady chain formation time. These present strong evidence for the sympathetic cooling enhanced crystallization process.

The wide statistical distribution presents a challenge for deterministic quantum information processing: individually addressing ions requires precise calibration of the tightly-focused laser positions~\cite{Lee2016}, which is difficult since the self-assembled Coulomb crystals change equilibrium positions and spacing for different ion numbers~\cite{James1998}. To obtain fine control over loaded ion number, we implement a conditional logic based on PMT counts and apply feedbacks. A scripted control sequence programmed in LabRAD software package~\footnote{AMO distribution, see for example M. Ramm thesis, UC Berkeley.} automatically reads the photon counts and controls the ablation laser and the RF trap potential. Fig.~\ref{fig:Control loading} (a) shows our control logic: if the PMT counts are lower than the threshold, we apply an extra loading attempt; if the counts are high than the threshold, we lower the RF potential by 13dB for 10 s to purposefully let some ions escape the trap. During this evaporation process, the trap is shallow and the cooling efficiency is reduced, thereby effectively ejecting undesirable hot ions. Such a conditional loading sequence yields a narrow distribution of ion number between 10 to 15 shown in Fig.~\ref{fig:Control loading}(b). We reduce the statistical fluctuation by a factor of 4.5 when comparing the standard deviations of the distribution and that of loading without feedback.

In addition to ion number control, ion crystals suitable for quantum information processing also need to be free of defects from contaminant ions. While direct ablation plasma is usually accompanied with ions from other isotopes or molecules, such problems are naturally eliminated for beryllium: $^{9}$Be is the only stable isotope, and its chemical properties give rise to a thin oxide layer that prevents the metal target from further reaction with air. In principle, we could load small amounts BeO$^{+}$, which would show up as dark ions with inter-ion spacing reduced by a factor of 1.4~\cite{James1998}. The change in equilibrium positions should be distinctly detectable on the camera images. However, this was never observed in our experiments. Instead, after continuous laser-cooling of a few minutes, we observe molecular ion (BeH$^{+}$) formation due to inelastic collisions with background hydrogen gas~\cite{Sawyer2015}. We apply the same low-RF evaporation process, allowing dark ions not directly interacting with the cooling laser to leave the trap. Fig.~\ref{fig:Control loading}(c) shows camera images before and after this sequence, a chain of 20 ions can be purified, i.e. evaporating 4 molecular ions and leaving 16 bright atomic ions. 




\color{black}


Finally, we observe an effect well-understood in earlier literature~\cite{Leibrandt2007}: the loading becomes less efficient when ablating on the same spot after a few dozen loading events. The ablation laser essentially micro-machines a crater into the target, making it more difficult for the plasma to escape. A picomotor-controlled mirror mount can be used to automatically randomize the ablation position for long-term consistent loading efficiency.

\section{Conclusion and Outlook}

In summary, we demonstrate a flexible and non-invasive loading scheme with feedback control for producing entropy-free beryllium linear ion chains. Our method presents several advantages over the existing thermal schemes and is easy implement. We uncover the non-Poisson statistics from sympathetically enhanced crystallization. Once upgraded with deterministic ion counting with a camera, we could remove the remaining fluctuations and obtain full control of the ion number. Demonstrating this in a macroscopic ion trap with only a few electrodes would provide a route to scalable quantum processors beyond 100 ions. Such stable and deterministic operation in macroscopic traps can be crucial for the next generation quantum computers in the NISQ eara and beyond~\cite{Preskill2018}.

\section{acknowledgments}
We thank Luis Orozco, Frank Moscatelli, Ross Flaxman, Umang Mishra, Yiheng Lin, and Ting-Rei Tan for helpful discussions and critical reading of the manuscript. This work was supported by the Department of Energy. 

\bibliography{ablation.bib}
\bibliographystyle{ieeetr}

\end{document}